\documentclass[print]{aa}

\usepackage{graphicx}

\usepackage{natbib}
\usepackage{txfonts}

\begin{document}

\title {A comparison of the s- and r-process element evolution in local dwarf spheroidal galaxies and in the Milky Way}
\titlerunning{S- and r-process evolution}

\author{Gustavo A. Lanfranchi\inst{1},
    Francesca  Matteucci\inst{2, 3},
\and    Gabriele Cescutti\inst{2}}

\institute{N\'ucleo de Astrof\'\i sica Te\'orica, Universidade
Cruzeiro do Sul, R. Galv\~ao Bueno 868, Liberdade, 01506-000, S\~ao Paulo, SP, Brazil
\and Dipartimento di Astronomia-Universit\'a di Trieste, Via G. B. 
Tiepolo 11, 34131 Trieste, Italy
\and  I.N.A.F. Osservatorio Astronomico di Trieste, via G.B. Tiepolo 
11, I-34131}

\date{Received xxxx/ Accepted xxxx}

\abstract{}
{We study the nucleosynthesis of several neutron capture elements (barium, europium, lanthanum, and yttrium) in local group dwarf spheroidal (dSph) galaxies and in the Milky Way by comparing the predictions of detailed chemical 
evolution models with the observed data. }
{We compare the evolution of [Ba/Fe], [Eu/Fe], [La/Fe], [Y/Fe], [Ba/Y], [Ba/Eu], [Y/Eu], and [La/Eu] observed in dSph galaxies and in our Galaxy with predictions of detailed chemical evolution models. The models for all dSph galaxies and for the Milky Way are able to reproduce several observational features of these galaxies, such as a series of abundance ratios and the stellar metallicities distributions. The Milky Way model adopts the two-infall scenario, whereas the most important features of the models for the dSph galaxies are the low star-formation rate and the occurrence of intense galactic winds. }
{ We predict that the [s-r/Fe] ratios in dSphs are generally different than the corresponding ratios in the Milky Way, at the same [Fe/H] values. This is interpreted as a consequence of the time-delay model coupled with different star formation histories. In particular, the star-formation is less efficient in dSphs than in our Galaxy and it is influenced by strong galactic winds. Our predictions are in very good agreement with the available observational data.}
{ The time-delay model for the galactic chemical enrichment coupled with different histories of star formation in different galaxies allow us to succesfully interpret the observed differences  in the abundance ratios of s- and r- process elements, as well as of $\alpha$-elements in dSphs and in the Milky Way. These differences strongly suggest that the main stellar populations of these galaxies could not have had  a common origin and, consequently, that the progenitors of local dSphs might not be the same objects as the building blocks of our Galaxy.
}

 \maketitle

\section{Introduction}

Are the formation and evolution of the Milky Way (MW) linked to its neighbouring dwarf galaxies, in the sense that physical processes (such as star-formation or galactic winds) occurring in one type of galaxy influenced the physical processes occurring in the other type? This a question that has been asked several times recently (Venn et al. 2004; Helmi et al. 2006; Geisler et al. 2007 and references therein) and a few clues have emerged, but no definitive answers have given. The Local Group is an excellent laboratory for shedding some light on that matter, due to the possibility of studying the properties of different types of galaxy in details. This includes large disk galaxies (such as the Milky Way) and dwarfs, in particular irregulars and spheroidals (Mateo 1998; Kunth $\&$ Ostlin 2000). 

The knowledge of how local galaxies formed and evolved provides important clues to the formation of structures in the universe and the whole subject of galaxy formation. The two most popular scenarios for the assembly of galaxies contradict each other, because the monolithical scenario proposes that large galaxies formed through the central collapse of a gas cloud at high redshifts (Larson 1974; Sandage 1986; Arimoto $\&$ Yoshii 1987; Matteucci 1994; Calura, Matteucci $\&$ Menci 2004), whereas the hierarchical  cosmological scenario states that large galaxies formed via the merger of minor blocks in more recent epochs (Sommerville et al. 2001; Menci et al. 2002). According to the second scenario, the dwarf spheroidal galaxies (dSph) of the local group could be the remains of the blocks that gave rise to our galaxy.
Therefore, our Galaxy and local dSph galaxies should share some common evolutionary features and properties, especially chemical signatures. A similar scenario was originally proposed by Searle \& Zinn (1978) for the formation of the outer halo of the Milky Way.

One possible approach to searching for evidence of this joint evolution is to analyse the chemical history of stars in the Milky Way and in its satellite dSph galaxies (Venn et al. 2004). By adopting this approuch, Geisler et al. (2007) analysed the scenario in which the Halo was formed by the same progenitors as the local dSph galaxies. They found no evidence of this when comparing the chemical properties of stars in the halo of the Milky Way and in local dSph galaxies, in particular the Sculptor galaxy.

The evolution of chemical elements is normally studied by analysing the abundance ratios of certain key elements leading to a better understanding of galaxy evolution (Tinsley 1980, Matteucci 1996). Comparing the observed patterns of these abundance ratios with predictions from chemical evolution models helps constrain several parameters of the models, thereby providing clues to understanding important processes  in the formation and evolution of galaxies and in the nucleosynthesis of chemical elements.

Abundance ratios predicted by the models, unlike other observables, strongly depend only on the adopted nucleosynthesis prescriptions, initial mass function (IMF), and star-formation rate (SFR). Consequently, comparing between models predictions and observations allows one to either constrain both the IMF and SFR of a galaxy if the nucleosynthesis of the elements analysed is well known or to obtain clues to the formation of the elements observed if the IMF and SFR are established. 
As a consequence, differences in the patterns of the same abundance ratios in different types of galaxies, on the other hand, reflect the particularities in the evolution of each type of system (Matteucci 1996). A comparison between abundance patterns in dSphs and the Milky Way therefore is necessary to ascertain wether these galaxies had a common origin or a completely different history of formation and evolution.

In this work we analyse the trends of Ba and Eu, plus Y and La in the MW and in six local dSph galaxies (Carina, Draco, Sagittarius, Sculptor, Sextan, and Ursa Minor). By comparing the observed trends in these galaxies with chemical evolution models suited to adjusting their main features, we are able to impose constraints on the production of these elements and also to shed some light on the possible evolutionary connection between the Milky Way and its neighbouring dSph galaxies. A comparison between the chemical evolution of the Milky Way and dSphs uses the standard model developed by Lanfranchi \& Matteucci (2003, hereafter LM03).
Previous papers (LM03, Lanfranchi \& Matteucci 2004 (LM04)) already studied the
[$\alpha$/Fe] ratios in dSphs and predicted that they are generally lower than the corresponding ratios in the Milky Way
at the same [Fe/H]. This is because the star formation in dSphs was assumed to have been much less efficient than in our Galaxy, thus producing the appearance of Fe from Type I a supernovae (SNe) at lower [Fe/H] values than in the Milky way.

The paper is organised as follows. In Sect. 2 we present
the observational data concerning the dSph galaxies and the Milky Way; in Sect. 3 we summarise the main characteristics of the adopted chemical evolution models for both types of galaxies, such as the star formation, the galactic winds and nucleosynthesis prescriptions; in Sect. 4 the predictions of our models are 
compared to observed abundance ratios and the results discussed.
Finally in Sect. 5 we draw some conclusions. All elemental
abundances are normalised to the solar values
([X/H] =  log(X/H) - log(X/H)$_{\odot}$)  as measured by 
Grevesse $\&$ Sauval (1998).

\section{Data sample}

The abundance of several chemical species in stars of local dSph galaxies have become available in the past few years due to the advent of large telescopes and high-resolution spectrographs (such as the Very Large Telescope - VLT - the FLAMES, and others). With this class of instruments several observational groups (Bonifacio et al. 2000, 2004; Shetrone, Cot\'e $\&$ Sargent 2001; Shetrone et al. 2003; Venn et al. 2004; Sadakane et al. 2004; 
Fulbright, Rich $\&$ Castro, 2004; Geisler et al. 2005) have devoted an enormous effort to determining chemical abundances in red giant stars in a wide range of ages, making it possible to analyse the chemical enrichment history of these galaxies. From these observations we then gathered the data from the same galaxies in LM03 and LM04.
Despite the relativly small number of data points, it is possible to compare the observed abundance ratios with the model predictions. We compared the evolution of [Ba/Fe], [Y/Fe], [La/Fe], [Eu/Fe], [Ba/Eu], [La/Eu], [Ba/Y], and [Y/Eu] observed in local dSph galaxies and in the Milky Way with the predictions of
chemical evolution models.

Following Lanfranchi, Matteucci $\&$ Cescutti (2006a - LMC06a), we adopted the updated abundance values of Venn et al. (2004) for the dSph galaxies to have a more homogeneous sample with data coming from different authors and to make a more proper comparison with the predictions of the models. In LMC06a, however, we only analysed the elements Ba and Eu. We extend the analysis to Yttrium and Lanthanum, for which there  is a new set of data for Sagittarius. The abundances of La in Sagittarius, in contrast to Ba and Eu from Bonifacio et al. (2000), could have been underestimated due to the fact that HFS was not used by Sbordone et al. (2007 - private communication). In that case, the adopted values should be seen in the plots as upper limits.
Like LMC06a, we excluded some observed stars from the sample since they exhibit anomalous values for [Ba/H] or [Eu/H]. Two stars in Ursa Minor and two in Sculptor might have had their heavy element abundance affected by external factors. In Ursa Minor, the stars K and 199 (in Shetrone, Cot\'e $\&$ Sargent 2001) exhibit heavy-element abundance ratios enhanced compared to those typical of other dSph stars: the K star has an abundance pattern dominated by the s-process and was classified as a carbon star, while Ursa Minor 199 is dominated by the r-process (see also Sadakane et al. 2004). In Sculptor,
the stars Sculptor 982 (Geisler et al. 2005) and Sculptor H-400 (Shetrone et al. 2003) exhibit enhanced heavy-element abundances. Both inhomogeneous mixing of the SNe II (Shetrone et al. 2003) and enrichment by another star, which had already ejected 
its material in the ISM (Geisler et al. 2005), were invoked to explain the anomalous abundance patterns of the above stars. Since these stars do not exhibit any abundance pattern characterised only by the nucleosynthesis process occurring in the stellar interior, keeping them in the sample could lead to a meaningless  comparison with the model predictions and, as a consequence, to a misleading interpretation and to incorrect conclusions about the processes and the site of production of the heavy elements analysed. 

For the Milky Way, we adopted the data taken from several works covering a wide range of metallicities from very metal-poor stars ([Fe/H] down to -4.0 dex) to stars with solar metallicities. The [Ba/Eu] ratio was collected from Fran\c cois et al. (2006), Burris et al. (2000), Fulbright (2000), Mashonkina $\&$ Gehren (2000, 2001), Koch $\&$ Edvardsson (2002), Honda et al. (2004), and Ishimaru et al. (2004).
The other heavy element abundances were taken from Fran\c cois et al. (2006), Cowan et al. (2002), Burris et al. (2000), Johnson (2002), Pompeia et al. (2003), and McWilliam $\&$ Rich (1994), in the case of lanthanum, and from Fran\c cois et al. (2006),
Burris et al. (2000), McWilliam et al. (1995),
Fulbright (2000, 2002),  Mashonkina $\&$ Geheren (2001), Johnson (2002),
Nissen $\&$ Schuster (1997),  Prochaska et al. (2000),
Gratton $\&$ Sneden (1994), Edvardsson et al. (1993), Stephens $\&$
Boesgaard (2002), and Honda et al. (2004) in the case of yttrium.

\begin{table*}
\begin{center}\scriptsize  
\caption[]{Models for dSph galaxies.}
\begin{tabular}{lccccccc}  
\hline\hline\noalign{\smallskip}  
galaxy &$\nu(Gyr^{-1})$ &$w_i$
&n &t($Gyr$) &d($Gyr$) &$t_{GW}(Gyr)$ &$IMF$\\    
\noalign{\smallskip}  
\hline
Carina &0.15 &5 &4 &0/2/7/9 &2/2/2/2 &0.53 &Salpeter\\
Draco  &0.05 &4 &1 &0 &4 &1.97 &Salpeter\\
Sagittarius &3.0 &9 &1 &0 &13 &0.10 &Salpeter\\
Sculptor &0.2 &13 &1 &0 &7 &0.44 &Salpeter\\
Sextan &0.08 &9 &1 &0 &8 &0.78 &Salpeter\\
Ursa Minor &0.1 &10 &1 &0 &3 &0.43 &Salpeter\\
Standard &0.3 &10 &1 &0 &8 &0.33 &Salpeter\\
\hline\hline
\end{tabular}
\end{center}
\end{table*} 

\section{Models} 

In this work, we investigate the evolution of heavy elements in dSph galaxies and compare it to the evolution of the same elements in the Milky Way by means of detailed chemical evolution models, one suited to the dSph galaxies and the other to the Milky Way. The models for the dSph galaxies are the same as described in LM03 and LM04, whereas the model for the Milky Way is the one from Chiappini, Matteucci, Gratton (1997 - CMG97), as adopted by Cescutti et al. (2007). All models adopt up-to-date nucleosynthesis yields for intermediate-mass stars (IMS) and supernovae of both types (type Ia and type II) as well as the effects of these objects in the energetics of the interstellar medium (ISM). The models allow the evolution of the abundances of several chemical elements (such as H, He, C, N, O, Mg, Si, Ca, Fe, Ba, Eu, La, Y, Sr, Zr) to be followed in detail, starting from the matter reprocessed by stars and injected in the ISM through galactic winds and supernovae (SNe) explosions.

The time evolution of the abundance in mass fraction of an element $\it i$ in the gas of the galaxy is described by the basic equations given in Matteucci (1996) and Tinsley (1980). In particular, the variation in time of the fractional mass of a chemical element, $G_i$, is given by

\begin{equation}
\dot{G_{i}}=-\psi(t)X_{i}(t) + R_{i}(t) + (\dot{G_{i}})_{inf} -
(\dot{G_{i}})_{out}
\end{equation}

\noindent
where the gas mass in the form of an element $\it i$ normalised to a total fixed mass, $M_{tot}$, is $G_{i}(t)=M_{g}(t)X_{i}(t)/M_{tot}$ and the total fractional mass of gas present in the galaxy at the time t is $G(t)= M_{g}(t)/M_{tot}$. The abundance by mass of an element $\it i$ is represented by the quantity $X_{i}(t)=G_{i}(t)/G(t)$, with the summation over all elements in the gas mixture being equal to unity. The star formation rate (SFR), namely the fractional amount of gas turning into stars per unit time, is represented by $\psi(t)$, and specified by the star formation efficiency $\nu$, namely  the inverse of the SF time-scale and expressed in $Gyr^{-1}$ . Here, $R_{i}(t)$ represents the returned fraction of matter in the form of an element $\it i$ that the stars eject into the ISM through stellar winds and supernova explosions, a term that contains all the prescriptions concerning the stellar yields and 
the supernova progenitor models. The infall of external gas and galactic winds are accounted for by the terms 
$(\dot{G_{i}})_{inf}$ and $(\dot{G_{i}})_{out}$, respectively. The type Ia SN progenitors are assumed to be white dwarfs in binary systems according to the formalism originally developed by Greggio $\&$ Renzini (1983) and Matteucci $\&$ Greggio (1986).

\subsection{The chemical evolution model for dSph galaxies}

In our scenario the dSph galaxies are formed through the infall of pristine gas until a mass of $\sim 10^8 M_{\odot}$ is accumulated. The star formation history
(SFH) of each galaxy is given by the analysis of observed colour-magnitude diagrams (CMDs), which suggest a unique long ($t \sim 3$ to $t \sim 8 \;Gyr$) episode of SF in five systems, in particular, Draco, Sagittarius, Sculptor, Sextan, and Ursa Minor (Hernandez et al. 2000; Aparicio et al. 2001;  Carrera et al. 2002; Dolphin et al. 2005). In the case of Carina, on the other hand, the SFH is characterised by four long ($\sim$ 2 Gyr of duration) episodes, as suggested in Rizzi et al. (2003), and successfully adopted in Lanfranchi, Matteucci $\&$ Cescutti (2006b - LMC06b). The episodes of SF, in our models, are strongly affected by the occurrence of galactic winds. As soon as the wind starts, a large fraction of the gas reservoir that fuels the SF is removed from the galaxy, causing a sudden drop in the star formation rate. In that sense, the majority of galactic stars 
are formed before the galactic wind occurs, characterising this epoch as the most important period of galactic activity.

The occurrence of the galactic wind depends on the binding energy of the galaxy and on the thermal energy of the gas, in the sense that
when the second equates or it is larger than the first a galactic wind is deflagrated (see for example Matteucci
$\&$ Tornamb\'e 1987). The thermal energy strongly depends on the assumptions for the thermalisation efficiency of supernovae of both types and of stellar winds (Bradamante et al. 1998). In this work we assume, as standard values, the thermalisation efficiencies suggested by Recchi et al. (2001), i.e. $\eta_{SNeIa} = 1.0$, $\eta_{SNeII} = 0.03$, $\eta_{SW} = 0.03$ (for SNeIa, SNeII and stellar winds, respectively). The binding energy, on the other hand, is strongly influenced by assumptions concerning the presence and distribution of dark matter (Matteucci 1992). A diffuse ($R_e/R_d$=0.1, where $R_e$ is the effective radius of the galaxy and $R_d$ is the radius of the dark matter core), but massive ($M_{dark}/M_{Lum}=10$) dark halo has been assumed for each galaxy. This particular configuration allows the development of a galactic wind in these small systems without destroying them. 

The rate of gas loss
{\bf $\dot G_{iw}$}, is proportional to the SFR, $\psi(t)$, through a proportionality constant, $w_i$, which is a free
parameter describing the efficiency of the galactic wind:

\begin{eqnarray}
 \dot G_{iw}\,=\,w_{i} \, \psi(t).
\end{eqnarray}

\noindent
The values adopted for the efficiency of the wind ($w_i$) are high (from 4 to 13) in order to reproduce the intense decrease observed in several abundance ratios, to remove a large fraction of the gas content of the galaxy, and to explain the observed stellar metallicity distributions (see Lanfranchi $\&$ Matteucci 2007 for a more detailed discussion). The comparison with all these constraints leaves little room for modifying the wind efficiency without changing the previous results much.

The main assumptions of the dSph galaxy models are

\begin{itemize}

\item
the model is one zone with instantaneous and complete mixing of gas inside
this zone;

\item
no instantaneous recycling approximation is adopted, i.e. the stellar 
lifetimes are taken into account;

\item
the evolution of several chemical elements (H, D, He, C, N, O, 
Mg, Si, S, Ca, Fe, Ba, Eu, La, Y, Sr and Zr) is followed in detail;

\item
the nucleosynthesis prescriptions include the yields of 
Nomoto et al. (1997) for type Ia supernovae, Woosley $\&$
Weaver (1995) (with the corrections suggested by 
Fran\c cois et al. 2004) for massive stars ($M > 10 M_{\odot}$), 
van den Hoek $\&$ Groenewegen (1997) for intermediate mass stars 
(IMS), and the ones described in Cescutti et al. (2006, 2007) and Busso et al. (2001)  for Ba and Eu.
\end{itemize}

The prescriptions for the SF (which follow a Schmidt law - Schmidt 1963), initial mass function (IMF - Salpeter 1955), infall, and galactic winds are the same as in LM03 and LM04. The main parameters adopted for the model of each galaxy, together with the predicted time for a galactic wind, $t_{GW}$, can be seen in Table 1, where $\nu$ is the star-formation efficiency, $w_i$ the wind efficiency,  $n$, $t$, and $d$ are the number, time of occurrence, and duration of the SF episodes, respectively, $t_{GW}$ the time of the galactic wind.

\subsection{The chemical evolution model for the Milky Way}

We adopted the two-infall model (CMG97) for the evolution of the Galaxy. In this scenario, the Galaxy formed through two main episodes of infall: the first episode formed the halo and thick disk on a timescale of $\sim$ 0.7 Gyr and the second one formed the thin disk on a timescale of $\sim$ 7 Gyr for the solar vicinity. The formation of the thin disk is assumed to be a function of the galactocentric distance, leading to an inside-out build-up of the Galactic disk (see also Matteucci $\&$ Fran\c cois 1989). Several concentrical rings, 2 kpc wide, with no exchange of matter between them, simulate the Galactic disk. Two important characteristics of this model are (i) the almost independent evolution of the halo and the thin disk and (ii) the assumed threshold in the SF. When the gas density decreases below a critical value, in particular 7 $M_{\odot}\;pc^{-2}$, the SF is halted (Kennicutt 1989, 1998; Martin $\&$ Kennicutt 2001).

This model is able to reproduce a number of observational constraints of the solar vicinity and the whole Galaxy. The evolution of several elements as a function of [Fe/H] (such as C, N, $\alpha$-elements, iron peak elements), several abundance ratios, the abundance gradients, and the G-dwarf metallicity distribution can be listed as the most important ones. In particular, Cescutti et al. (2007) used the same model to study the abundance gradients of several chemical elements including $\alpha$-elements, iron peak elements, and heavy elements such as Ba, Eu, and La.

The rate of mass accretion A(r,t) (the infall rate) is a function of time and galactocentric distance and is given by

\begin{equation}
A(r,t)=a(r)e^{-t/\tau_{H}}+b(r)e^{(t-t_{max})/\tau_{D}(r)}
\end{equation}

\noindent
where $t_{max}=1Gyr$ is the time for the maximum infall rate on the thin disk and 
 $\tau_{H}=0.8Gyr$ and $\tau_{D}$ are, respectively, the time scale for the formation of the halo thick-disk and the timescale of the thin disk. This last quantity is a function of the galactocentric distance:
\begin{equation}
\tau_{D}=1.033r(kpc)-1.267Gyr.
\end{equation}

The coefficients $a(r)$ and $b(r)$ are constrained to reproduce the present-day total surface mass density as a function of galactocentric distance. In particular, $b(r)$ is assumed to be different from zero only for $t>t_{max}$, where $t_{max}$ is the time of maximum infall on the thin disk (see Chiappini et al. 2003, for details).

The adopted law for the SFR is substantially a Schmidt law with a dependence also on the total surface mass density. In particular,

\begin{equation}
\psi(r,t)=\nu\left(\frac{\Sigma(r,t)}{\Sigma(r_{\odot},t)}\right)^{2(k-1)}
\left(\frac{\Sigma(r,t_{Gal})}{\Sigma(r,t)}\right)^{k-1}G^{k}_{gas}(r,t),
\end{equation}

where the efficiency of the star formation process, $\nu$, is set to be $1Gyr^{-1}$ for the disk ($t\ge1Gyr$).
The total surface mass density is represented by $\Sigma(r,t)$, whereas $\Sigma(r_{\odot},t)$ is the total surface mass density at the 
solar position. The expression $G_{gas}(r,t)$ represents the surface density normalised to the present time
total surface mass density in the disk $\Sigma_{D}(r,t_{Gal})$, where $t_{Gal}=13.7Gyr$ is the age 
assumed for the Milky Way and $r_{\odot}=8kpc$ the solar galactocentric distance 
(Reid 1993). The exponent of the surface gas density, $k$, is set equal to 1.5.
These choice of values for the parameters allow the model to fit the observational constraints very well, in particular in the solar vicinity.
We recall that, below a critical threshold for the gas surface density 
($7M_{\odot}pc^{-2}$ for the thin disk and $4M_{\odot}pc^{-2}$ for the halo phase), 
we assume that the star formation is halted. The adopted IMF is the one proposed by Scalo (1986).

\begin{table*}
\caption{The stellar yields for La and Y in massive stars (r-process)
in the case of primary origin.} \label{rLa}
\centering
\begin{minipage}{90mm}
\begin{tabular}{|c|c|c|}
\hline
$M_{star}$  & $ X_{La}^{new}$ & $ X_{Y}^{new}$\\
\hline\hline
12.   & 9.00$\cdot10^{-8}$ & 4.00$\cdot10^{-7}$\\ 
15.   & 3.00$\cdot10^{-9}$ & 1.00$\cdot10^{-8}$\\   
30.   & 1.00$\cdot10^{-10}$ & 1.00$\cdot10^{-9}$\\ 
\hline\hline
\end{tabular}
\end{minipage}
\end{table*}

\section{Nucleosynthesis prescriptions}

The adopted nucleosynthesis prescriptions are the same as in previous works (see Lanfranchi, Matteucci $\&$ Cescutti 2006a - LMC06a) and in Cescutti et al. (2007). In particular, the prescriptions for the Fe and $\alpha$-elements (namely O, Si, Ca, Mg) are those suggested in Fran\c cois et al. (2004). These authors analysed several theoretical stellar yields by comparing of observed [el/Fe] vs. [Fe/H] trends with predictions for the Milky Way model in the solar neighbourhood and selected the sets of yields that fit the data best. 
For single low-intermediate mass stars, the stellar yields are those from van den Hoek $\&$ Groenewegen (1997)
for the case of the mass loss parameter that varies with metallicity (see Chiappini et al. 2003, model5), whereas the yields of SNe Ia are taken from 
Iwamoto et al.(1997) with revisions in the case of Mg. The yields of SNe II that they suggest provides a best fit are the Woosley $\&$ Weaver (1995) ones. In this case, no modifications are required for the yields of Ca and Fe as computed for solar composition. For oxygen the best results 
are given by the Woosley $\&$ Weaver (1995) yields computed as functions of the metallicity. For the other elements, variations in the predicted yields are required to best fit the data (see Fran\c cois et al. 2004 for details).

In the case of neutron process elements (s- and r-process), we adopted the yields of Cescutti et al. (2006) and LMC06a for Ba and Eu and the yields of Cescutti et al. (2007) for La. In these works, the yields of s-process elements (in particular La, Ba, and Y) are those predicted by Busso et al. (2001) in the mass range $1.5-3M_{\odot}$, with an extension to the mass range $1.5-1M_{\odot}$. The r-production of Ba and Eu were obtained by following the prescriptions of model 1 of Cescutti et al. (2006). These yields are empirical and were chosen to provide the best fit to the abundances of Ba and Eu in low-metallicity stars in the Milky Way, as measured by Fran\c cois et al. (2007). The same set of yields were also adopted by LMC06a in the comparison between the predictions of the dSph galaxy models with the abundances of Ba and Eu in red giant stars in local dSph resulting in a good agreement.

Barium is assumed to be mainly produced by s-process in low and intermediate-mass stars (LIMS) in the mass range $1 - 3 M_{\odot}$ with a minor fraction also produced as an r-process element in massive stars (12 - 30 $M_{\odot}$). Europium on the other hand is considered 
to be a pure r-process element produced in massive stars in the same range of masses. For La, we followed the prescriptions of Cescutti et al. (2007), who adopted the same prescriptions of Ba: a major contribution from s-process in LIMS and a minor production as r-process in massive stars in the same mass range as Ba. The yields for the s-process are taken from Busso et al. (2001), whereas those from r-process are the ones from Cescutti et al. (2007) (see Table 2).

\section{Results}

It has been suggested that Local Group dSph galaxies could be the remaining blocks of the small systems that gave rise to our Galaxy (White $\&$ Rees 1978; Navarro, Frenk $\&$ White 1997).  From the chemical evolution point of view, one can analyse such possible connection by comparing the abundance and abundance ratios of several elements and the metal-poor tail of the stellar metallicity distributions observed in the Milky Way and in local dSphs (Venn et al. 2004; Ripamonti et al. 2007; Helmi et al. 2006, Geisler et al. 2007).  It is expected, for example, that the s- and r-process ratios in dSph galaxies might be different from the same ratios in stars with comparable metallicities in our Galaxy, due to  different SFHs. In the dSphs the SFR is much lower, so the contribution from the s-process should appear at lower metallicities when compared to the Milky Way, because of the slower increase in [Fe/H] with time.
Venn et al. (2004) examined the abundance and abundance ratios of several s-process and r-process elements in dSph and in the Milky Way stars and claim that no significant stellar component of our Galaxy appears to have been formed from the dSphs. Helmi et al. (2006) uses a completely different approach to reach the same conclusion. They compared the metal-poor tail of the stellar metallicity distributions of four dSph galaxies to the one of the Galactic halo and found significant differences. The differences, according to the authors, suggest that the progenitors of local dSph galaxies might have been fundamentally different from the building blocks that formed our Galaxy.

We intend not only to examine such a subject but also to investigate the different evolutions of heavy elements in these galaxies in order to constraint the formation and evolution of these elements and the evolution of these dwarf galaxies. By means of the detailed chemical evolution models described above,
we are able to predict  the evolution of La, Ba, Y, and Eu.
The predictions for the dSphs allow us to fill the gaps in the observations, given the reduced number of available data. At high or very low metallicities, for example, there is almost no observed data for the dSphs, which prevents a comparison to extend to the whole range of metallicities observed in our Galaxy.

\begin{figure}
\centering
\includegraphics[height=9cm,width=9cm]{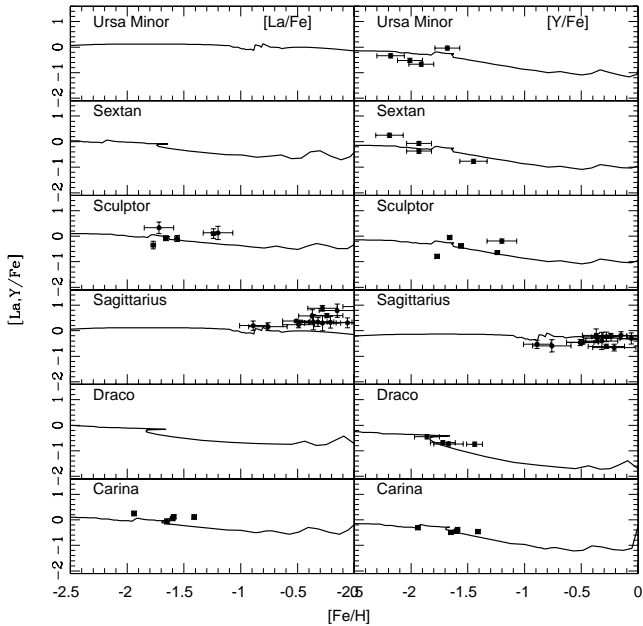}
\caption[]{The evolution of [La/Fe] and [Y/Fe] predicted by the dSph model for all six galaxies.
} 
\end{figure}

For the Milky Way we applied the two-infall model (CMG97) as adopted by Cescutti et al. (2006). In the case of the dSph, galaxies we adopted the standard model of LM03. This model was constructed to reproduce the observed [$\alpha$/Fe] ratios of a sample of dSph galaxies (Carina, Draco, Fornax, LeoI, Leo II, Sagittarius, Sculptor, Sextan, and Ursa Minor) as a whole, without separating each individual galaxy. We consider that the general trends of the heavy elements in the dSph galaxies are the same for all individual galaxies, in spite of the observed differences caused by the particularities in the evolution of each galaxy (in particular, different SF efficiencies and galactic wind efficiencies). By doing that, we intend to highlight the differences between the pattern of these elements in the Milky Way and its spheroidal satellites. There are, consequently, a few points that lie above or below the model's prediction, depending on the characteristics of the galaxy (SF history, SFR, galactic wind efficiency). Since the aim of the standard model is to draw a general trend for all analysed dSph galaxies , small offsets between the predictions from the standard model and the data of individual galaxies should not be seriously taken into account. The goal is to highlight the differences between the Milky Way and the dSph galaxies as a whole.

\begin{figure}
\centering
\includegraphics[height=8cm,width=8cm]{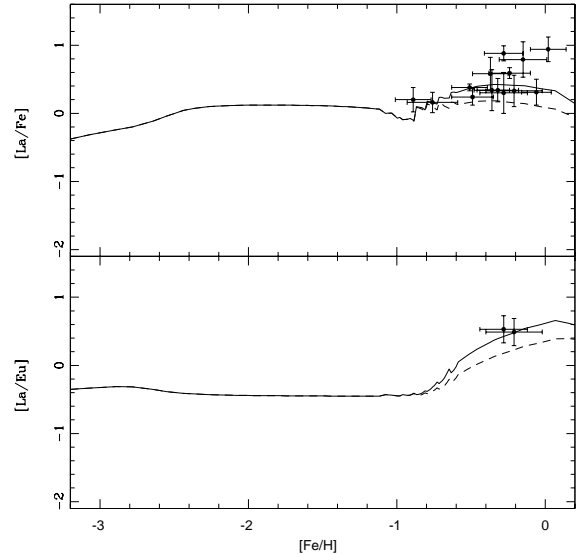}
\caption[]{The evolution of [La/Fe] and [La/Eu] predicted by the Sagittarius model with La yields in LIMS increased by a factor of 2 and 4 (dashed and solid lines, respectively). We ruled out the model with the La yields increased by a factor of 4, even though it provides a better fit to the dSph observations, because it cannot reproduce the Milky Way data.} 
\end{figure}

We should mention, however, that Sagittarius dSph galaxy should be seen as an exception, not only for the particular chemical characteristics (elemental abundances, abundance ratios - Sbordone 2006 - star formation efficiency, and metallicity distribution - LM04), but also for the fact that this galaxy is clearly 
interacting with our Galaxy (Mateo et al. 1996; Majewski et al. 2003.) The other dSph galaxies exhibit similar chemical properties (similar values for the chemical abundances, abundance ratios, SF efficiencies - LM04) with small differences due to their particular formation and evolution. The same properties in Sagittarius, on the other hand, are more like our Galaxy than like other dwarfs (see also Geisler et al. 2007). Consequently, when comparing local dSph galaxies with the Milky Way galaxy, we decided to exclude Sagittarius.

\begin{figure*}
\centering
\includegraphics[height=10cm,width=10cm]{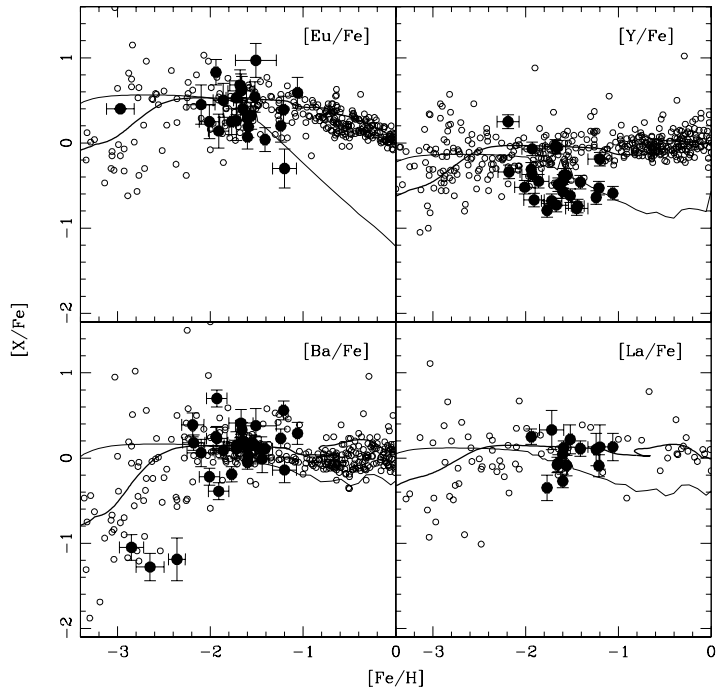}
\caption[]{The evolution of Ba, La, Y, and Eu in the Milky Way and local dSph galaxies. The thick line corresponds to the predictions of the Milky Way model and the thin line the predictions of the standard model for the dSph galaxies.}
\end{figure*}

In Figure 1 we show the predictions of La and Y in the 6 dSphs.
There is no available data for La in Draco, Ursa Minor, and Sextan, so we only show the model predictions in these cases. 
The predicted and observed trends of these two elements can be explained by the s- and r-process contributions from stars in different mass ranges and by the occurrence of strong galactic winds. At low metallicities, the r-production in massive stars plays a dominant role by being responsible for the plateau at almost solar values for [La/Fe] and [Y/Fe]. At intermediate to high metallicities ([Fe/H] $>$ -1.8 dex), the injection of these elements in the ISM by LIMS stars becomes important in the dSph galaxies. Besides that, at [Fe/H] $\sim$ -1.6 dex, a galactic wind is developed, thereby removing a large fraction of the gas content of the galaxy and reducing the SFR considerably. With a low SFR, the contribution from massive stars is almost negligible, whereas the one from LIMS remains important. These combined effects result in a decrease in the abundance ratios.

The difference between the trends among these elements are related to details of their nucleosynthesis, in particular the different amounts of the fractional mass produced by each process for each element. The fraction of Y produced in LIMS by s-process is lower than the one of La. Consequently, one can observe a very similar trend of [La/Fe] and [Y/Fe] in all galaxies, but with a more intense decrease in [Y/Fe] after the occurrence of galactic winds, due to its lower production in LIMS. 

In all six dSph galaxies we are able to satisfactorily reproduce the observed data for all elements, maybe with the exception of La in Sagittarius. In this case, there is a discrepancy between model predictions and observations, in the sense that we under-predict the values of [La/Fe] at high metallicities (see Figure 1). It seems, in fact, that there is a trend in the observed data of increasing [La/Fe] with increasing metallicity, whereas in the model predictions one notices a slight decrease. The decrease is a result of the effects of galactic winds on the SFR, as mentioned before. Could this discrepancy be solved with a change in the value of the galactic wind efficiency? Probably not. If we change the efficiency of the wind for Sagittarius, we would lose the agreemeent with other abundance ratios, not only with the [alpha/Fe] ratios but also with other heavy elements. From Figure 1, it is clear that if we increase the wind efficiency (it would be necessary to match La data), the agreement with [Y/Fe] would be lost. In that case, we should look for another explanation for this discrepancy.
In that metallicity range, the production of La is almost totally dominated by the s-process in LIMS. Consequently,  an under-prediction of La in LIMS could be the source of the discrepancy. We adopted the yields of Busso et al. (2001) for La in this mass range, which could be increased by hand to get a better fit to the observations. In that case, however, we would face a serious problem. The same set of yields when adopted in the model of the Milky Way allows very good agreement with the observed data, including the solar value (see Cescutti et al. 2006). Consequently, if we change the yields considerably (for a factor of 2 or more) we would lose the agreement with the Milky Way data and the solar value.

If the yields cannot be considerably modified, the other source of discrepancy could reside in the data. The data for La in Sagittarius at high metallicity comes entirely from the work of Sbordone et al. (2006). In such work the authors did not use hyper-fine splitting (HFS) in the calculation of La abundances, which could cause an over-prediction of the suggested values, but not as high as the difference between the observed data and the model predictions  (Sbordone, private communication). In that sense, the combined effects of overestimating the observed values with the possible under-prediction of La yields in LIMS could be the cause of the discrepancy between the observed data and model predictions. In order to examine this scenario, we ran two alternative models in which the La yields in LIMS were increased by a factor of two and four (Figure 2). One can notice that an increase in the yields of La in LIMS by a factor of two is not enough to reach the observed high values, which are only fitted by the model with yields four times higher than the ones of Busso et al. (2001). Since the agreement obtained with the solar neighbourhood and the solar values is lost if we adopt yields increased more than two times, then it is highly probable that there is also a overestimation of the observed values. 

The comparison between the predictions of the models and the observed values for neutron capture elements strongly suggests that La and Y trends can be explained in all dSph galaxies (with maybe the exception of Sagittarius - see discussion above) by models that adopt a scenario in which both these elements are produced by the s-process in LIMS, in the mass range $M = 1-3M_{\odot}$, with a minor fraction produced by the r-process in massive stars ($M = 12-30M_{\odot}$), following closely the production of Ba. Strong galactic winds also play a fundamental role in explaining the behaviour of La and Y at intermediate to high metallicities, as in the case of Ba, Eu, and $\alpha$ elements.

\begin{figure*}
\centering
\includegraphics[height=10cm,width=10cm]{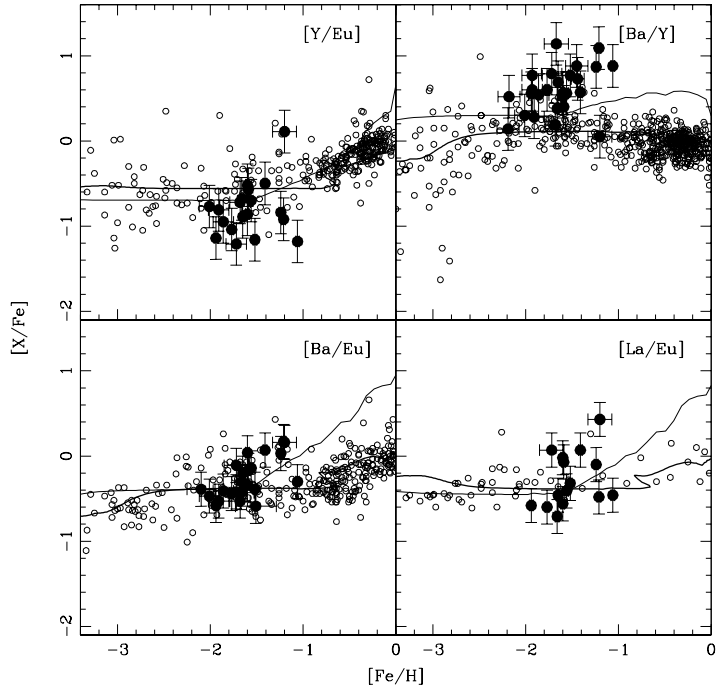}
\caption[]{The evolution of [Ba/Y], [Ba/Eu], [La/Eu], and [Y/Eu] in the Milky Way and local dSph galaxies. The thick line corresponds to the predictions of the Milky Way model and the thin line to the predictions of the dSph galaxies standard model.  }
\end{figure*}

In Figure 3, we show the predictions of the models of both dSphs and the Milky Way  for Ba, La, Y, and Eu along with the observed data for the studied systems (small open circles: Milky Way stars, large symbols: dSph galaxies).
This model was constructed to reproduce the observed [$\alpha$/Fe] ratios of a sample of dSph galaxies (Carina, Draco, Fornax, LeoI, Leo II, Sagittarius, Sculptor, Sextan, and Ursa Minor) as a whole, without separating each individual galaxy, so this model  represents a typical dSph galaxy.

At a first glance, it seems that the data from the two types of galaxies overlap at intermediate metallicities, which could suggest a common origin. A more careful analysis, however, reveals that there are marked differences. In both types of galaxies, the observed values show a large dispersion at low metallicities, which makes the interpretation more complex. This wide spread is normally attributed to an inhomogeneous mixing; i.e., the abundance pattern of r-process elements of a newly-formed star was contaminated by SNe II explosion in the vicinity. Even with the dispersion, one notices that, while the abundances of La, Ba, and Eu are similar in the Milky Way and in the dSph galaxies, [Y/Fe] exhibits significant lower values in the majority of the stars in dSph galaxies when compared to the ones of the Milky Way. This offset in [Y/Fe] suggests that the enrichment from r- and s-process elements in the two types of galaxies might have been different, as already suggested by Venn et al. (2004). Besides that, the predictions of the models suggest meaningful differences at low and, especially, at intermediate-to-high metallicities. 

\begin{figure*}
\centering
\includegraphics[height=10cm,width=10cm]{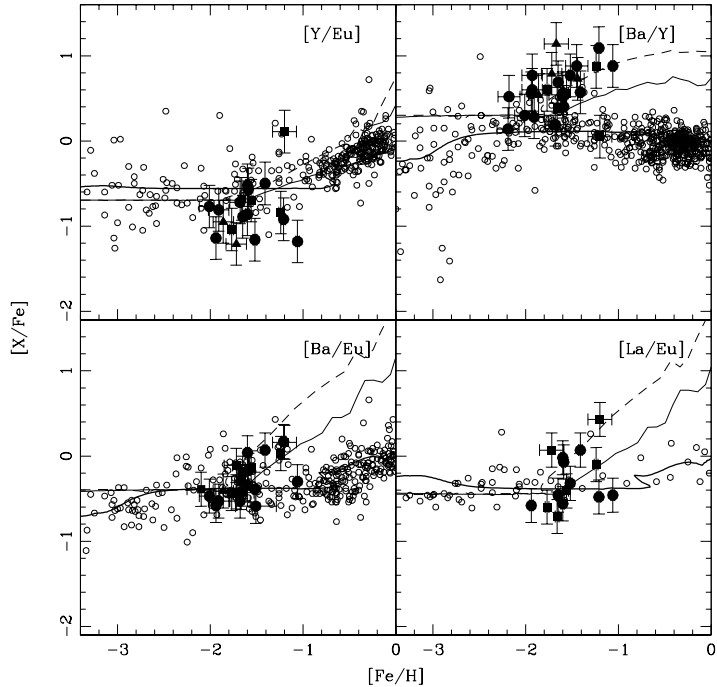}
\caption[]{Comparison between the evolution of [Ba/Y], [Ba/Eu], [La/Eu], and [Y/Eu] predicted by the Milky Way model and the Draco and Sculptor models. The thick solid line corresponds to the predictions of the Milky Way model, the thin solid line to the Sculptor model, and the thin dashed line to the Draco model. The data from Sculptor are represented by solid squares, the data from Draco by solid triangles, and all the other dSphs by the solid circles.}
\end{figure*}

At low metallicities ([Fe/H] $\sim$ -2.5 dex), there are several stars with low abundances of [La/Fe], [Ba/Fe] and [Eu/Fe] in the Milky Way, whereas these elements exhibit solar values in the dSph. There are a few stars in the dSph that show low values of [Ba/Fe], similar to the ones in the Milky Way, but they could not be reproduced by the model with the same nucleosynthesis prescriptions adopted in the model of our Galaxy. Since these stars belong to different galaxies (Ursa Minor, Draco, and Sextans) and come from different works (Venn et al.  2004, who homogenized the atomic data for spectral lines, and Sadakane et al. 2004), it is not likely that the low [Ba/Fe] values are representative of one specific galaxy or that they are caused by a systematic error. In that way, these low values, if confirmed by further observations, could only be reproduced if a different production of Ba is adopted. In such a case, the r-production of Ba in massive stars should be ignored and the s-production should be extended to stars with masses up to $4 M_{\odot}$ (see LMC06a). In either case, the metal-poor stars that enriched the ISM of the dwarfs and the MW could not be the same, suggesting that these metal-poor stellar populations are different (see also Helmi et al. 2006). If one considers the predictions of the models, the two populations are characterised by different abundance patterns, but if one considers the stars observed, then the production of heavy elements is not the same. It should be noticed, however, that the numbers of metal-poor stars observed in local dSphs are still small and further observations are required to confirm the low values of [Ba/Fe]. These two facts suggest that if metal-poor stars in local dSph galaxies do exist with low abundances of neutron capture elements, similar to the values observed in our Galaxy, then a different mechanism for the production of r- and s-process elements in dSph galaxies should be invoked.

From intermediate-to-high metallicities, there are almost no stars observed in local dSphs, but the predicted trends are quite different from the one in our Galaxy. In the dSphs, the abundances of neutron capture elements (as all other elements) are strongly influenced by the development of strong galactic winds. This decreases the abundance of r-products but not as much as the s-products, giving rise to very low [Eu/Fe]. In the Milky Way, the feature that plays a major role in the abundance patterns of neutron capture elements at high metallicity is the s-production of these elements by LIMS. Since these stars have a long lifetime (up to a few Gyrs), they only contribute to the enrichment of the ISM when [Fe/H] has had time to increase to intermediate or high values, depending on the SFR. In the Milky Way, the injection of s-products from LIMS marked by the change in the slope of the [Ba/Eu] ratio both in the data and in models predictions) starts becoming important at around [Fe/H] $\sim$ -1.0 dex, whereas in local dSphs, it happens at lower metallicities, around [Fe/H] $\sim$ -1.7 dex. The reason for this difference in the model can be explained by the fact that, even though Ba is injected into the ISM on the same timescale in both types of galaxies, the dSphs are less enriched in [Fe/H] than the ISM in the Milky Way at a particular time, owing to the less efficient star formation assumed in the dSphs compared to to the Galaxy. This is an important result that derives directly from the time-delay model (delayed Fe production relative to alpha-elements) coupled with different SFHs (see Matteucci 2003). The changes in the abundance patterns of heavy elements at different metallicities in the Milky Way and in the dSph, due to the injection of fresh elements by LIMS into the ISM, also strongly suggest that the two metal-poor populations on the different types of galaxies are not the same and do not have a common origin.

In Figure 4, we show the predictions of the models of both galaxies for [Ba/Y], [Ba/Eu], [La/Eu], and [Y/Eu], along with the observed data for the studied systems. 
The analysis of abundance ratios between heavy elements strengthens the above scenario. At high metallicities one can notice an increasing trend of [Ba/Eu], [La/Eu], and [Y/Eu] in dSphs much more intense than the ones observed in the Milky Way. In our Galaxy, the increase in [Ba/Eu] is normally attributed to contributions of the s-process from LISM, whereas in dSph the interruption of the enrichment of r-process elements from massive stars due to the galactic winds plays a major role. The ratio [Ba/Y] reflects the production of light and heavy s-process elements, since Ba is almost totally produced by s-process and the fraction of Y produced by s-process is smaller. In the Milky Way, the [Ba/Y] values are always lower than in the dSph galaxies, indicating different productions of these two elements in both types of galaxies. In the same way, several stars in the dSph galaxies exhibit lower [Y/Eu] abundance ratios when compared to stars with the same metallicity in our Galaxy. The nucleosynthesis of r- and s-process elements also seems to have followed different tracks in each type of galaxy, as is seen in the analysis of [Y/Eu]. From low-to-intermediate metallicities, [Y/Eu] is lower in the dSph galaxies than in the MW, suggesting different amounts of r- and s-process elements in each type of galaxy.

In Figure 4, one sees that the prediction of the standard model for dSph for [Ba/Y] does not reach the highest values observed. As mentioned before, this model was adjusted to fit the data of all dSph galaxies at the same time, leaving behind the differences (in particular $\nu_i$ and $w_i$) among the systems. The stars with the highest values of [Ba/Y] belong to Draco, galaxy that is characterised by a much lower SF efficiency ($\nu_i$ = 0.05) and by a lower wind efficiency ($w_i$ = 4). These values allow the specific model for Draco to reproduce the highest [Ba/Y], which are not fitted by the standard model (Figure 5). In Figure 5, we also show the predictions of the Sculptor dSph galaxy. By comparing the models for individual galaxies with the model for the Milky Way, it can be seen the differences are even more pronounced. In particular, the differences between the predictions from both Draco and Sculptor models and the Milky Way model are even larger (when compared to the standard model) for [Ba/Y], [Ba/Eu], and [La/Eu] at high metallicities.

Several differences in the evolutionary tracks of Ba, La, Eu, and Y, in the MW and local dSph galaxies are noticed in the comparison of their evolution from both the observational point of view and model predictions. The much lower SFR and the intense galactic winds in local dSph galaxies are the main features responsibles for the marked differences at low ([Fe/H] $<$ -2.0 dex) and high ([Fe/H $>$ -1.6 dex) metallicities. Consequently, we suggest that the very old, metal-poor, stellar populations present in these galaxies are not the same and do not share a common origin.

\section{Summary}

We analysed the evolution of several r- and s-process elements (in particular barium, europium, lanthanum, and yttrium) in Local Group dwarf spheroidal (dSph) galaxies and in the Milky Way by comparing observed data in a wide range of metallicities with predictions of detailed chemical evolution models. The adopted models for all dSph galaxies and for the Milky Way are the ones described in LM03 and Cescutti et al. (2006), respectively. The Milky Way model adopts the two-infall scenario, whereas the most important features of the models for the dSph galaxies are the low star formation rate and intense galactic winds. By assuming that Eu is a pure r-process elements synthesised in massive stars ($M = 12-30M_\odot$), whereas Ba, La, and Y are produced mainly by s-process in low and 
intermediate-mass stars (LIMS) in the mass range $M = 1-3M_\odot$ with a minor contribution from r-process in massive stars, we are able to reproduce the observed abundance trends of all abundance ratios very well. 
At low metallicities the trends are defined by the contribution from the r-process in massive stars in both galaxies, whereas at high metallicities the s-process in LIMS are the main contributor to the enrichment of these elements in our Galaxy and in the dSph galaxies. In these last ones, however, strong galactic winds play a fundamental role at [Fe/H] $>$ -1.6 dex. When comparing the data and predictions of the Milky Way model with the standard model for the dSph galaxies, significant differences both in the observed trend and in models predictions at all metallicities are seen. These differences strongly suggest that the metal-poor stellar populations are not the same in dSphs and the Milky Way.

The main conclusions are the following:

\begin{itemize}

\item
The observed trends of [La/Fe] and [Y/Fe] are reproduced very well by the models of all six dSph galaxies and by the Milky Way model, suggesting that our assumed nucleosynthesis prescriptions are reasonable. In particular, we assume that La and Y are are mainly s-process elements produced in LIMS ($M = 1 -3 M_\odot$) with a minor fraction produced by r-process in massive stars in the range $M = 12-30M_\odot$.

\item
The Sagittarius model seems to underpredict the values of [La/Fe] at high metallicities when compared to observations. One possible source of this disagreement could be an underestimation of La yields in LIMS by Busso et al. (2001). A model with an increased yield of La by a factor of two can reproduce the data very well if they are overestimated by no more than $\sim$ 0.2 dex. An increase greater than a factor of two would, however, cause a disagreement with the Milky Way data, so it is not likely that this is the only source of disagreement.

\item
The predictions for [Ba/Fe], [Eu/Fe], [La/Fe], and [Y/Fe] by the dSph standard model developed by LM03 and the Milky Way model reveals several differences, as suggested by observed data. At low metallicities ([Fe/H] $<$ -2.0 dex), the models of dSph galaxies predict higher values for these ratios due to the lower SFR, whereas) they are always lower at high metallicities ([Fe/H] $>$ -1.4 dex, due to the effects of galactic winds on the SFR. At intermediate metallicities,  [Ba/Fe], [Eu/Fe], and [La/Fe] seem to be similar in both types of galaxies but [Y/Fe] is lower in the dSph galaxies.

\item 
The heavy element abundance ratios ([Y/Eu], [Ba/Eu], [La/Eu], and [Ba/Y]) also exhibit different trends in our Galaxy and its neighbour dwarfs: at high metallicities, the values in dSph galaxies are always higher, whereas at intermediate metallicities [Y/Eu] is lower and [Ba/Y] is higher in the dSphs.

\item 
There are a few dSph stars with very low [Ba/Fe] at low metallicities, similar to what is observed in stars with comparable metallicities in our Galaxy. If these values are indeed real, a different nucleosynthesis of Ba in these galaxies is required. In such a case, the r-production of Ba in massive stars should be suppressed and the s-production should be extended to stars with $M = 4 M_\odot$ (see LMC06a).

\item 
The meaningful differences between the trends of heavy elements in local dSph galaxies and our Galaxy strongly suggest that the stellar populations of these two types of galaxies are different and might not have shared a common evolution. In fact, the evolution in the dSph galaxies must have proceeded at a much lower rate than in the Milky Way and been strongly influenced by intense galactic winds. In that sense, it is unlikly that the progenitors of local dSph galaxies and the building blocks of the Milky Way are the same objects.

\end{itemize}

\section*{Acknowledgments}
G.A.L. acknowledges financial support from the Brazilian agency 
FAPESP (proj. 06/57824-1). F.M. acknowledges financial support
from  the Italian  I.N.A.F. (Italian National Institute for
Astrophysics), project PRIN-INAF-2005-1.06.08.16


\begin{thebibliography}{}

\bibitem{}
Aparicio, A., Carrera, R., Mart\'\i nez-Delgado, D., 2001, AJ, 122, 2524

\bibitem{}
Arimoto N., Yoshii Y., 1987, A\&A, 173, 23 

\bibitem{}
Bonifacio P., Hill V., Molaro P., Pasquini L., Di Marcantonio P.,
Santin P., 2000, A\&A, 359, 663 

\bibitem{}
Bonifacio P., Sbordone L., Marconi G., Pasquini L., Hill V., 2004, 
A\&A, 414, 503

\bibitem{}
Bradamante, F., Matteucci, F., D'Ercole, A., 1998, A\&A, 337, 338
  
\bibitem{} 
Burris D.L., Pilachowski C.A., Armandroff T.E., 
Sneden C., Cowan J.J., Roe H., 2000, ApJ, 544, 302

\bibitem{} 
Busso M., Gallino R., Lambert D.L., 
Travaglio C., Smith V.V.,
2001, ApJ, 557, 802

\bibitem{}
Calura F., Matteucci F., Menci N., 2004, MNRAS, 353, 500

\bibitem{}
Carrera, R., Aparicio, A., Martnez-Delgado, D., Alonso-Garca, J., 2002, AJ, 123, 3199

\bibitem{}
Cescutti, G., Fran\c cois, P., Matteucci, F., Cayrel, R., Spite, M., 2006, A\&A,
448, 557

\bibitem{}
Cescutti, G., Matteucci, F., Fran\c cois, P., Chiappini, C., 2007, A\&A, 
462, 943

\bibitem{}
Chiappini C., Matteucci F., Gratton R. 1997, ApJ, 477, 765 

\bibitem{}
Chiappini C., Romano D., Matteucci F., 2003, MNRAS, 339, 63

\bibitem{}
Cowan J.J., Sneden C., Burles S., 
Ivans I.I., Beers T.C., Truran J.W., 
Lawler J.E., Primas F., Fuller G.M., 
Pfeiffer B., Kratz K-L.,  2002, ApJ, 572, 861


\bibitem{} 
Dolphin, A.E., Weisz, D.R., Skillman, E.D., Holtzman, J.A., 2005, 
to appear in Valls-Gabuad D. \& Chavez M., eds.,
Resolved Stellar Populations, ASP Conference Series, astro-ph/0506430

\bibitem{} 
Edvardsson B., Gustafsson B., Johansson S. G., Kiselman D., Lambert D. L., Nissen P. E., Gilmore G., 1994, A\&A, 290, 176

\bibitem{}
Fran\c cois, P., Matteucci, F., Cayrel, R., 
Spite, M., Spite, F., Chiappini, C., 2004, A\&A, 421, 613

\bibitem{}
Fran\c cois, P., Depagne, E., Hill, V., et al., 2006, AIPC, 847, 205

\bibitem{}
Fran\c cois, P., Depagne, E., Hill, V., et al., 2007,  A\&A, 476, 935

\bibitem{}
Fulbright J.P., 2000, AJ, 120, 1841

\bibitem{}
Fulbright J.P., 2002, AJ, 123, 404

\bibitem{}
Fulbright, J.P., Rich R.M., Castro S., 2004, ApJ, 612, 447

\bibitem{}
Geisler D., Smith V.V., Wallerstein G., 
Gonzalez G., Charbonnel C., 2005, AJ, 129, 1428

\bibitem{}
Geisler D., Wallerstein G., Smith V.V., Casetti-Dinescu D. I., 2007, PASP, 119, 939

\bibitem{}
Gratton R.G., Sneden C., 1994, A\&A, 287, 927

\bibitem{}
Greggio, L., Renzini, A., 1983, A\&A, 118, 217

\bibitem{} 
Grevesse N., Sauval A.J., 1998, Space Science Reviews, 85, 161 

\bibitem{} 
Helmi A., Irwin M.J., Tolstoy E., 2006, ApJ, 651L, 121

\bibitem{}
Hernandez X., Gilmore G., Valls-Gabaud D., 2000, MNRAS, 317, 831

\bibitem{}
Honda S., Aoki W., Kajino T. et al., 2004, ApJ, 607, 474

\bibitem{}
Johnson J.A., 2002, ApJS, 139, 219

\bibitem{}
Ishimaru Y., Wanajo S., Aoki, W., Ryan, S.G., 2004, ApJ, 600L, 47

\bibitem{}
Iwamoto, K., Brachwitz F., Nomoto K., Kishimoto N., Umeda H., Hix W.R., Thielemann
F.-K., 1999, ApJS, 125, 439

\bibitem{}
Kennicutt R.C.Jr., 1989, ApJ, 344, 685

\bibitem{}
Kennicutt R.C.Jr., 1998, ApJ, 498, 541

\bibitem{}
Koch A., Edvardsson B., 2002, A\&A,381, 500

\bibitem{}
Kunth, D., Ostlin, G., 2000, A\&ARv, 10, 1

\bibitem{}
Lanfranchi, G., \& Matteucci, F., 2003, MNRAS, 345, 71
 
\bibitem{}
Lanfranchi, G., \& Matteucci, F., 2004, MNRAS, 351, 1338
 
\bibitem{}
Lanfranchi, G., Matteucci, F., \& Cescutti, G., 2006a, MNRAS, 365, 477

\bibitem{}
Lanfranchi, G., Matteucci, F., \& Cescutti, G., 2006b, A\&A, 453, 67

\bibitem{}
Lanfranchi, G., \& Matteucci, F., 2007, A\&A, 468, 927

\bibitem{}
Larson, R. B., 1974, MNRAS, 166, 585

\bibitem{}
Majewski, S. R.; Skrutskie, M. F.; Weinberg, M. D., Ostheimer, J. C., 2003, ApJ, 599, 1082

\bibitem{}
Martin, C.L., Kennicutt, R.C.Jr., 2001, ApJ, 555, 301

\bibitem{}
Mashonkina, L., Gehren, T., 2000, A\&A, 364, 249

\bibitem{}
Mashonkina, L., Gehren, T., 2001, A\&A , 376, 232

\bibitem{}
Mateo, M.L., 1998, ARA\&A, 36, 435

\bibitem{}
Mateo, M., Mirabal, N., Udalski, A., Szymanski, M., Kaluzny, J., Kubiak, M., Krzeminski, W., Stanek, K. Z., 1996, ApJ, 458L, 13

\bibitem{}
Matteucci, F., 1992, ApJ, 397, 32

\bibitem{}
Matteucci F., 1994, A\&A, 288, 57

\bibitem{}
Matteucci, F., 1996, FCPh, 17, 283

\bibitem{}
Matteucci, F. 2003, Ap\&SS, Vol.284, p.539

\bibitem{} 
Matteucci, F.,\&  Greggio, L., 1986, A\&A, 154, 279
Reid M.J., 1993, ARAA, 31, 345
\bibitem{} 
Matteucci, \& F., Tornamb\'{e}, A., 1987, A\&A, 185, 51

\bibitem{}
Matteucci, F., Fran\c cois, P., 1989, MNRAS, 239, 885

\bibitem{}
Menci, N., Cavaliere, A., Fontana, A., Giallongo, E., Poli F.,
2002, ApJ, 575, 18

\bibitem{} 
McWilliam, A., Preston, G.W., Sneden, C., Searle, L., 1995, AJ, 109, 2757

\bibitem{} 
McWilliam, A., Rich R. M., 1994, ApJS, 91, 749Stephens A., Boesgaard A.M., 2002, AJ, 123, 1647

\bibitem{}
Navarro, J. F., Frenk, C. S., White, Simon D. M., 1997, ApJ, 490, 493

\bibitem{} 
Nissen, P.E., Schuster, W.J., 1997, A\&A, 326, 751

\bibitem{} 
Nomoto, K., Hashimoto, M., Tsujimoto, T. et al., 1997, 
Nucl. Phys. A, 616, 79

\bibitem{} 
Pompeia, L., Barbuy, B., Grenon, M., 2003, ApJ, 592, 1173

\bibitem{} 
Prochaska, J.X., Naumov, S.O., Carney, B.W., McWilliam, A., Wolfe, A.M., 2000, AJ,
120, 2513

\bibitem{} 
Recchi, S., Matteucci, F., D'Ercole, A., 2001, MNRAS, 322, 8

\bibitem{}
Reid, M.J., 1993, ARAA, 31, 345

\bibitem{}
Ripamonti, E., Tolstoy, E., Helmi, A., Battaglia, G., Abel, T., 2007, EAS, 24, 15

\bibitem{} 
Rizzi, L., Held, E.V., Bertelli, G., Saviane, I., 2003, ApJL, 589, 85

\bibitem{}
Sadakane, K., Arimoto, N., Ikuta, C., Aoki W.,
Jablonka, P., Tajitsu, A., 2004, PASJ, 56, 1041

\bibitem{} 
Salpeter, E.E., 1955, ApJ, 121, 161

\bibitem{}
Sandage, A., 1986, A\&A, 161, 89

\bibitem{}
Sbordone, L., Bonifacio, P., Buonanno, R., Marconi, G., Monaco, L., Zaggia, S., 2006, A\&A, 464, 201 

\bibitem{} 
Scalo, J.M., 1986, FCPh, 11, 1

\bibitem{}
Schmidt, M., 1963, ApJ, 137, 758

\bibitem{}
Searle, L., Zinn, R., 1978, ApJ, 225, 357

\bibitem{}
Shetrone, M., C\^ot\'e, P., \& Sargent, W.L.W., 2001, 
ApJ, 548, 59

\bibitem{}
Shetrone, M., Venn, K.A., Tolstoy, E., Primas, F., 2003, AJ, 125, 684

\bibitem{}
Somerville, R. S., Primack, J. R., Faber, S. M., 2001, MNRAS,
320, 504

\bibitem{}
Stephens A., Boesgaard A.M., 2002, AJ, 123, 1647

\bibitem{} 
Tinsley B.M., 1980, 1980, FCPh ,5, 287

\bibitem{} 
van den Hoek L.B., Groenewegen M.A.T. 1997, A\&A Suppl., 123, 305

\bibitem{}
Venn K. A., Irwin M., Shetrone M.D. et al., 2004, AJ, 128, 1177

\bibitem{}
White S. D. M., Rees M. J., 1978, MNRAS, 183, 341

\bibitem{}
Woosley S.E., Weaver T.A., 1995, ApJS, 101, 181  

\end{thebibliography}
\end{document}